\RequirePackage{ifpdf}
\documentclass[12pt,a4paper]{JHEP3}

\usepackage{amssymb, amsmath, amsopn, amsthm, dsfont}
\usepackage{epsfig}
\usepackage{graphicx}
\usepackage{caption}
\usepackage{subcaption}

\allowdisplaybreaks

\newcommand{\cN}{\mathcal{N}}

\newcommand{\g}{\gamma}

\newcommand{\cn}{{\cal N}}

\newcommand{\co}{{\cal O}}

\newcommand{\cw}{{\cal W}}

\newcommand{\nn}{\nonumber}

\def\bal#1\eal{\begin{align}#1\end{align}}
\def\alp[#1]{\begin{align}#1\end{align}}

\newcommand{\beq}{\begin{equation}}
\newcommand{\eeq}{\end{equation}}

\newcommand{\e}{\epsilon}

\newcommand{\p}{\partial}

\newcommand{\be}{\begin{equation}}
\newcommand{\ee}{\end{equation}}
\newcommand{\bea}{\begin{eqnarray}}
\newcommand{\eea}{\end{eqnarray}}
\newcommand{\bean}{\begin{eqnarray*}}
\newcommand{\eean}{\end{eqnarray*}}

\def\>{\rangle}
\def\<{\langle}
\def\a{\alpha}

\def\e{\epsilon}

\def\g{\gamma}

\def\p{\pi}

\def\F{\Phi}

\def\L{\Lambda}

\def\Q{\Theta}

\def\pa{\partial}
\newcommand{\fn}{\footnotemark\footnotetext}

\title{ Higgsing the stringy higher spin symmetry}

\author{Matthias R.\ Gaberdiel${}^{\diamondsuit}$, Cheng Peng${}^{\diamondsuit}$, and Ida G. Zadeh${}^{\spadesuit}$\\

{\tt gaberdiel@itp.phys.ethz.ch, pengch@itp.phys.ethz.ch, zadeh@brandeis.edu}\\
${}^\diamondsuit$ Institut f\"ur Theoretische Physik, ETH Zurich,
CH-8093 Z\"urich, Switzerland \\
${}^{\spadesuit}$ Martin Fisher School of Physics, Brandeis University, Waltham, MA 02454, USA
}

\abstract{It has recently been argued that the symmetric orbifold theory of $\mathbb{T}^4$ is dual to 
string theory on AdS$_3\times {\rm S}^3 \times \mathbb{T}^4$ at the tensionless point. At this point in moduli space, 
the theory possesses a very large symmetry algebra that includes, in particular, a ${\cal W}_\infty$ algebra 
capturing the gauge fields of a dual higher spin theory. Using conformal perturbation theory, we study the behaviour of 
the symmetry generators of the symmetric orbifold theory under the deformation that corresponds to switching on the string tension. 
We show that the generators fall nicely into Regge trajectories, with the higher spin fields
corresponding to the leading Regge trajectory. We also estimate the form of the Regge trajectories for large 
spin, and find evidence for the familiar logarithmic behaviour, thereby suggesting that the symmetric orbifold theory is 
dual to an AdS background with pure RR flux.}

\preprint{BRX-TH-6297}

\allowdisplaybreaks

\begin{document}
\tableofcontents
\section{Introduction}

Recently some progress relating the higher spin/CFT duality to the stringy
AdS/CFT correspondence was made for the case of AdS$_3$. In particular, it was shown in \cite{Gaberdiel:2014cha} 
that a certain limit of the CFT dual of the higher spin theory 
with $\cn=4$ supersymmetry \cite{Gaberdiel:2013vva} forms a closed subsector of the symmetric orbifold
theory, the CFT dual of string theory on ${\rm AdS}_3 \times {\rm S}^3 \times \mathbb{T}^4$. 

Among other things, this result suggests that the symmetric orbifold theory $(\mathbb{T}^{4})^{\otimes N} / S_N$ is dual to 
string theory at the tensionless point. At this point in moduli space, the symmetry algebra of string theory is much bigger 
than just the ${\cN=4}$ 
superconformal algebra, and indeed even bigger than the vector-like symmetry algebra of the Vasiliev higher spin theory
\cite{Vasiliev:2003ev,Prokushkin:1998bq,Prokushkin:1998vn}. In fact, the usual picture one has in mind is that the 
Vasiliev fields correspond to the leading
Regge trajectory that become massless (higher spin fields) at the tensionless point. By the same token, one should
then expect that also the higher Regge trajectories lead to massless higher spin fields, and this seems indeed to be in line 
with the structure of the dual CFT, as studied in detail in \cite{Gaberdiel:2015mra}. 

In this paper we want to confirm, at least qualitatively, this picture by studying the perturbation of the symmetric orbifold
theory that corresponds to switching on the string tension. Under this perturbation, we expect that the symmetry algebra is
broken down to the $\cN=4$ superconformal algebra. In addition, 
we should expect that the masses of the fields associated 
to the Vasiliev ${\cal W}_\infty$ algebra should be smallest, at each given spin, relative to those of the additional higher spin fields
(that correspond to the higher Regge trajectories). Furthermore, one may hope that we see something like `Regge' trajectories
to emerge in this regime. Using conformal perturbation theory we will find what we believe to be very convincing evidence 
for this picture: to the extent to which
we have managed to determine the anomalous dimensions, those of the original ${\cal W}_\infty$ algebra are indeed smallest
(at a given spin). Furthermore, the additional symmetry generators of the stringy symmetry algebra seem to fall naturally
into different Regge trajectories, where the $n^{\rm th}$ trajectory corresponds to the symmetrised polynomials of $n+1$
free fields of the symmetric orbifold theory. We also find evidence that 
the anomalous dimensions $\gamma(s)$ behave as $\log s$ at large spin $s$, at least before any mixing effects are taken into
account. This is in agreement with expectations from
the analysis of classical string solutions in AdS$_5$, see \cite{Gubser:2002tv} and, e.g., \cite{Beisert:2010jr} for a review, 
or the explicit results for AdS$_3$, 
see in particular \cite{Loewy:2002gf,David:2014qta}. 
\medskip

The paper is organised as follows. In Section~\ref{sec:hsf} we explain the structure of the higher spin fields of the symmetric orbifold. 
Section~\ref{sec:marginal} describes the exactly marginal operators that include  the moduli of the underlying
$\mathbb{T}^4$ torus, as well as the modulus from the $2$-cycle twisted sector 
that corresponds to switching on the string tension. In Section~\ref{fstpert} we show,
by a first order calculation, that the former do not break the higher spin symmetry, whereas the latter does.
In order to determine the anomalous dimensions (and hence the induced masses) quantitatively, we then turn to a second order
analysis in Section~\ref{sndpert}. We also explain the relation between anomalous dimensions and bulk masses there. 
In Section~\ref{sec:exp} we then describe the results of our explicit computations: in Section~\ref{sec:lowspin} we 
give the exact anomalous dimensions for the low spin operators to lowest non-trivial order in the perturbation, while in 
Section~\ref{sec:partialdiag} we calculate the diagonal matrix elements for the quadratic (leading Regge trajectory)
and cubic (first subleading Regge trajectory) generators at large spin. We discuss our results in Section~\ref{sec:discuss}. 
There are two appendices, where we give the explicit expressions for the higher spin generators in terms of the free fields
(Appendix~\ref{ffall}), and the integrals of the second order analysis of the perturbation calculation (Appendix~\ref{evintegral}). 
We have also included an ancillary file in the {\tt arXiv} submission of this paper  which contains the numerical values of the diagonal 
elements of the mixing matrix for the quadratic and cubic generators, as well as analytic expressions for these diagonal elements.

\section{The symmetric orbifold}\label{sec:hsf}

A duality between matrix extended higher spin theories on AdS$_3$ with large ${\cn=4}$ supersymmetry and the 
Wolf space coset models  was proposed in \cite{Gaberdiel:2013vva}, see also 
\cite{Creutzig:2013tja,Candu:2013fta,Creutzig:2014ula,Candu:2014yva}  for further developments of matrix extended 
higher spin theories on AdS$_3$.  In the limit where the large $\cN=4$ superconformal algebra contracts to the more
familiar small $\cN=4$ algebra (and hence the higher spin theory should correspond to 
${\rm AdS}_3 \times {\rm S}^3 \times \mathbb{T}^4$), the `perturbative part' of the Wolf space cosets becomes a closed
subsector of the symmetric orbifold $(\mathbb{T}^4)^{\otimes (N+1)} / S_{N+1}$, which in turn is thought to be dual to 
string theory on ${\rm AdS}_3 \times {\rm S}^3 \times \mathbb{T}^4$ at the tensionless point. This therefore ties in nicely 
with the general belief  that the perturbative higher  spin theory should describe a subsector of 
string theory in the tensionless limit.

In this section we want to describe the symmetry algebra of the symmetric or\-bi\-fold (that contains in particular
a ${\cal W}_{\infty}[0]$ algebra as a subalgebra).
Before applying the orbifold
projection the theory is generated by $2(N+1)$ complex bosons and $2(N+1)$ complex fermions, as well as their hermitian conjugates;
we shall denote these fields by
$\phi^{(i)}_a$, $\bar{\phi}^{(i)}_a$, $\psi^{(i)}_a$, $\bar{\psi}^{(i)}_a$, where  $i=1,2$, and $a=1,\ldots, N+1$ labels the different copies. 
(Here the $\bar{\phi}^{(i)}_a$ are the hermitian conjugates of the $\phi^{(i)}_a$, and similarly for the fermions.) Their non-trivial OPEs then 
have the form
\alp[
&\partial\bar{\phi}^{(i)}_a(z)\, {\partial\phi}^{(j)}_{a'}(w) \sim \frac{\delta^{ij}\delta_{aa'}}{(z-w)^2} + {\cal O}(1)\ ,\\
&\bar{\psi}^{(i)}_a(z)\, {\psi}^{(j)}_{a'}(w) \sim \frac{\delta^{ij}\delta_{aa'}}{(z-w)} + {\cal O}(1) \ .
]
The orbifold group $S_{N+1}$ acts by permuting the $N+1$ copies (labelled by $a$). The higher spin fields come from the untwisted 
sector of the orbifold, and the single-particle generators are all of the form
\begin{equation}\label{hs}
\sum_{a=1}^{N+1} P^{j_1}_a \cdots P^{j_m}_a \ , 
\end{equation}
where each $P^j_a$ is one of the $4$ bosons or fermions in the $a$'th copy, including possibly derivatives. In this language, the higher 
spin fields that  are dual to the Vasiliev theory are associated to the bilinear generators, i.e., to the fields (\ref{hs}) with $m=2$.
Strictly speaking the original ${\cal W}_{\infty}[0]$ algebra of \cite{Gaberdiel:2013vva} corresponds only to the neutral bilinear generators, 
i.e., to those where we have one unbarred and one barred generator. However, it is not difficult to see (see also 
Section~\ref{sec:quadratic} below) that if we extend
${\cal W}_{\infty}[0]$ by the generators corresponding to $(0;[2,0,\ldots,0])$ and $(0;[0,\ldots,0,2])$, 
we still obtain a Vasiliev-type higher spin algebra \cite{Gaberdiel:2015mra}.

The other generators of the stringy algebra can be organised 
in representations of this extended ${\cal W}_\infty$ algebra --- the columns of the Higher Spin Square \cite{Gaberdiel:2015mra} --- 
where we have one
column for each $m=2,3,4,\ldots$, with $m=2$ corresponding to the bilinear ${\cal W}_\infty$ generators.
It is then natural to believe that the higher spin fields associated to $m=2$ lie on the leading
Regge trajectory, those associated to $m=3$ on the first sub-leading Regge trajectory, etc. Indeed, 
in flat space, the different Regge trajectories are constructed, roughly speaking, by different numbers of creation operators,
and this is directly parallel to the above.   

\subsection{The quadratic fields}\label{sec:quadratic}

For the following it will be important to understand the structure of the single particle generators of the stringy algebra 
in a little more detail.
The original ${\cal W}_{\infty}[0]$ algebra of \cite{Gaberdiel:2013vva} 
is generated by the ${\cal N}=4$ superconformal algebra together with a number of multiplets,
\begin{equation}\label{Winf}
{\cal W}_{\infty}[0] \ : \qquad 
\bigl( {\cal N}=4 \bigr) \ \oplus \ \bigoplus_{s=1}^{\infty} R^{(s)} \ , 
\end{equation}
where $R^{(s)}$ is the ${\cal N}=4$ supermultiplet consisting of the generators
\begin{equation}
\begin{array}{lcc}\label{D2mult2}
& s: \quad & ({\bf 1},{\bf 1})  \\
& s+\tfrac{1}{2}: \quad& ({\bf 2},{\bf 2})  \\
R^{(s)}: \qquad & s+1: \quad & ({\bf 3},{\bf 1}) \oplus ({\bf 1},{\bf 3})  \\
& s+\tfrac{3}{2}: \quad & ({\bf 2},{\bf 2})  \\
& s+2: \quad & \  \  ({\bf 1},{\bf 1})  \ .
\end{array}
\end{equation}
Here the quantum numbers are the dimensions with respect to the $\mathfrak{su}(2)_+ \oplus \mathfrak{su}(2)_-$ 
algebra (that is contained in the ${\cal N}=4$ superconformal algebra).\footnote{At $\lambda=0$ 
the  $\mathfrak{su}(2)_+$ is only a global algebra, and the corresponding current generators only form a
$\mathfrak{u}(1)^3$ algebra, that is described by three of the four $\mathfrak{u}(1)$ currents in (\ref{bs1}) below.}
To be more specific, the ${\cal N}=4$  superconformal
algebra contains $4$ spin $s=\frac{1}{2}$ fields, that are described by the orbifold invariant combinations
\bal
&f^{(i)}=\sum_{a=1}^{N+1} \psi^{(i)}_a\ ,  \qquad \bar{f}^{(i)}=\sum_{a=1}^{N+1} \bar{\psi}^{(i)}_a\ , \label{4f}
\eal
where the subscript $a$ represent the different copies in the symmetric product. At spin $s=1$, the ${\cal W}_\infty[0]$ algebra
contains $8$ generators --- $7$ are contained in the ${\cal N}=4$ superconformal algebra, while $V_0^1$
is the bottom component of $R^{(1)}$. Explicitly, the former are the $\mathfrak{su}(2)_-$ generators
\bal
&J^+ =- \sum_{a=1}^{N+1}  :{\psi}^{(2)}_a\bar{\psi}^{(1)}_a:\ , \qquad 
J^- =- \sum_{a=1}^{N+1}  :{\psi}^{(1)}_a\bar{\psi}^{(2)}_a:\ , \\
\nn & J^3 =-\frac{1}{2} \sum_{a=1}^{N+1} \, (  :{\psi}^{(1)}_a\bar{\psi}^{(1)}_a:-:{\psi}^{(2)}_a\bar{\psi}^{(2)}_a:)\ , 
\eal
as well as the super-descendants of (\ref{4f}), 
\begin{equation}\label{bs1}
b^{(i)} = \sum_{a=1}^{N+1} \partial\phi^{(i)}_a \ , \qquad \bar{b}^{(i)} =  \sum_{a=1}^{N+1}  \partial\bar{\phi}^{(i)}_a \ ,
\end{equation}
where $i=1,2$. On the other hand, the bottom component of $R^{(1)}$ can be identified with 
\begin{equation}
V^1_0 =-\sum_{a=1}^{N+1}(  :{\psi}^{(1)}_a\bar{\psi}^{(1)}_a:+:{\psi}^{(2)}_a\bar{\psi}^{(2)}_a:)\ .
\end{equation}
\smallskip

As was mentioned above, the stringy algebra contains additional bilinear operators in the free fields that 
correspond to the representations $(0;[2,0,\ldots,0])$ and $(0;[0,\ldots,0,2])$, see \cite{Gaberdiel:2014cha}. 
The relevant wedge characters (including chemical potentials for both $\mathfrak{su}(2)$ algebras) are 
\begin{eqnarray}
\chi^{({\rm wedge})}_{(0;[2,0,\ldots,0])}  = 
\chi^{({\rm wedge})}_{(0;[0,\ldots,0,2])} & = & \frac{q}{(1-q)(1-q^2)} \, 
(1 + y_1 y_2 q^{1/2}) \, (1 + y_1 y_2^{-1} q^{1/2}) \\
& & \qquad \qquad \qquad \quad \times  (1 + y_1^{-1} y_2 q^{1/2}) \, (1 + y_1^{-1} y_2^{-1} q^{1/2}) \ ,  \nonumber 
\end{eqnarray}
and hence the analysis of \cite{Gaberdiel:2014cha,Gaberdiel:2015mra} implies that the additional generators lie in the multiplets
\begin{equation}\label{addq}
(0;[2,0,\ldots,0]) \ : \quad \bigoplus_{r=1}^{\infty} R^{(2r-1)} \ ,
\end{equation}
and similarly for $(0;[0,0,\ldots,0,2])$. Thus there are two additional $R^{(s)}$ multiplets for each odd spin $s$ from
$(0;[2,0,\ldots,0])$ and $(0;[0,\ldots,0,2])$. For example, for spin $s=1$, the relevant generators are 
 \bal
& C_{2}=\sum_{a=1}^{N+1}  :{\psi}^{(1)}_a{\psi}^{(2)}_a:\ , \qquad
\bar{C}_{2}=\sum_{a=1}^{N+1}  :\bar{\psi}^{(1)}_a\bar{\psi}^{(2)}_a:\ .
\eal
We have also written out explicitly the fields at $s=\frac{3}{2}$ in appendix~\ref{ffall}.

\subsection{The cubic and higher order fields}

The other fields of the stringy chiral algebra involve symmetrised higher order polynomials of the free fields. In particular,
it is natural to believe that the cubic fields, i.e., those associated to the ${\cal W}_\infty[0]$ representations
$(0;[n,0,\ldots,0,\bar{n}])$ with $m=n+\bar{n}=3$, lie on the first subleading Regge trajectory.
Using the analysis of  \cite{Gaberdiel:2015mra} we can read off the spin content of the relevant representations:
both $(0;[3,0,\ldots,0])$  and $(0;[0,\ldots,0,3])$  contribute each
\begin{equation}\label{2.13}
(0;[3,0,\ldots,0]) \ \hbox{and}  \ (0;[0,\ldots,0,3])  : \quad \bigoplus_{s=2}^{\infty} \, n(s)\,  \Bigl[ R^{(s)}({\bf 2},{\bf 1}) \
\oplus \ R^{(s+3/2)} ({\bf 1},{\bf 2})\Bigr] \ , 
\end{equation}
where the multiplicities $n(s)$ are determined by the generating function (see, e.g., Appendix~B of \cite{Gaberdiel:2013vva})
\begin{equation}\label{ns}
\frac{q^2}{(1-q^2)(1-q^3)} = \sum_{s=2}^{\infty} n(s) \, q^s  \ , 
\end{equation}
while the ${\cal N}=4$ multiplets $R^{(s)}({\bf 2},{\bf 1})$ and $R^{(s)}({\bf 1},{\bf 2})$
are of the form 
\begin{equation}
\begin{array}{lcc}\label{D2mult2+}
& s: \qquad & ({\bf 2},{\bf 1})  \\
& s+\tfrac{1}{2}:  \qquad& ({\bf 3},{\bf 2}) \oplus ({\bf 1},{\bf 2})  \\
R^{(s)}({\bf 2},{\bf 1}): \qquad & s+1: \qquad & ({\bf 4},{\bf 1}) \oplus ({\bf 2},{\bf 1}) \oplus ({\bf 2},{\bf 3})  \\
& s+\tfrac{3}{2}: \qquad & ({\bf 3},{\bf 2}) \oplus ({\bf 1},{\bf 2})   \\
& s+2: \qquad &   ({\bf 2},{\bf 1})  
\end{array}
\end{equation}
\begin{equation}
\begin{array}{lcc}\label{D2mult2-}
& s: \qquad & ({\bf 1},{\bf 2})  \\
& s+\tfrac{1}{2}:  \qquad& ({\bf 2},{\bf 3}) \oplus ({\bf 2},{\bf 1})  \\
R^{(s)}({\bf 1},{\bf 2}): \qquad & s+1: \qquad & ({\bf 1},{\bf 4}) \oplus ({\bf 1},{\bf 2}) \oplus ({\bf 3},{\bf 2})  \\
& s+\tfrac{3}{2}: \qquad & ({\bf 2},{\bf 3}) \oplus ({\bf 2},{\bf 1})  \\
& s+2: \qquad & ({\bf 1},{\bf 2}) \ .
\end{array}
\end{equation}
On the other hand, $(0;[2,0,\ldots,0,1])$  and $(0;[1,0,\ldots,0,2])$ contribute each
\begin{equation}\label{2.17}
(0;[2,0,\ldots,0,1]) \ \hbox{and}  \ (0;[1,0,\ldots,0,2])  : \quad \bigoplus_{s=3/2}^{\infty} \, m(s)\,  \Bigl[ R^{(s)}({\bf 1},{\bf 2}) \ \oplus \ 
R^{(s+1/2)}({\bf 2},{\bf 1})  \Bigr] \ , 
\end{equation}
where the multiplicities $m(s)$ are determined by the generating function
\begin{equation}\label{ms}
\frac{q^{3/2}}{(1-q)(1-q^2)} = \sum_{s=3/2}^{\infty} m(s) \, q^s  \ .
\end{equation}

The multiplicities of the higher order families of higher spin fields can be similarly determined. For example, for $m=4$, 
we get from $(0;[2,0,\ldots,0,2])$ the lowest order multiplets 
\begin{equation}\label{2.19}
(0;[2,0,\ldots,0,2]) : \quad R^{(2)} \  \oplus \ R^{(5/2)}({\bf 2},{\bf 2}) \ \oplus  \  R^{(3)} \ \oplus \ 
R^{(3)}({\bf 1},{\bf 3}) \ \oplus \ R^{(3)}({\bf 3},{\bf 1}) \ \oplus \ \cdots \ , 
\end{equation}
where $R^{(s)}({\bf d}_1 ,{\bf d}_2)$ is the ${\cal N}=4$ multiplet whose lowest component has spin $s$ and transforms in the 
$({\bf d}_1 ,{\bf d}_2)$ under the two $\mathfrak{su}(2)$ algebras.\footnote{Thus in this terminology,
$R^{(s)}\equiv R^{(s)}({\bf 1},{\bf 1})$. }
The other representations with $m=4$ start only at $s=5/2$ and $s=3$, respectively
\begin{equation}
(0;[3,0,\ldots,0,1]) \ \cong \ (0;[1,0,\ldots,0,3]) : \quad R^{(5/2)}({\bf 2},{\bf 2}) \ \oplus  \ R^{(3)} \ \oplus \ R^{(3)}({\bf 3},{\bf 1}) \ \oplus \cdots
\end{equation}
and
\begin{equation}
(0;[4,0,\ldots,0]) \ \cong \ (0;[0,\ldots,0,4]) : \qquad R^{(3)}({\bf 3},{\bf 1}) \ \oplus \ \cdots
\end{equation}
More generally, the fields that appear in the $m$'th column have spin $s\geq m-2$, and their number increases, for large spin $s$, 
as $s^{m-1}$.

\section{The Exactly Marginal Operators}\label{sec:marginal}

Next we need to identify the exactly marginal operators that induce the deformation of interest, i.e., that 
correspond to switching on the string tension. A priori, there are two types
of exactly marginal operators, see, e.g., \cite{David:2002wn}: those that come from the untwisted sector of the 
symmetric orbifold, and those that arise from the $2$-cycle twisted sector. The former correspond to the 
moduli that deform the shape and complex structure of the underlying torus, and hence should not break the
higher spin symmetry; on the other hand, the $2$-cycle twisted sector moduli deform the theory away from
the symmetric orbifold point and will turn out to break the higher spin symmetry. A certain linear combination 
of these deformations describes the perturbation that switches on the string tension.

In the following, we shall construct these moduli in terms of the free fields of the symmetric orbifold; we shall
then discuss the perturbation by them in turn. The exactly marginal operators that come from the untwisted sector are
\bal\label{upert}
&\nn \sum_{a=1}^{N+1} (\a^{(i)}_a)_{-1}(\tilde{\a}^{(j)}_a)_{-1}\big|0\>\ , \qquad
\sum_{a=1}^{N+1} (\bar{\a}^{(i)}_a)_{-1}(\tilde{\a}^{(j)}_a)_{-1}\big|0\>\ ,\nn\\
&\sum_{a=1}^{N+1} (\a^{(i)}_a)_{-1}(\tilde{\bar{\a}}^{(j)}_a)_{-1}\big|0\>\ , \qquad
\sum_{a=1}^{N+1} (\bar{\a}^{(i)}_a)_{-1}(\tilde{\bar{\a}}^{(j)}_a)_{-1}\big|0\>\ ,
\eal
where $(\alpha^{(i)}_a)_n$ and $(\bar\alpha^{(i)}_a)_n$ are the modes associated to $\partial \phi^{(i)}_a$ and
$\partial\bar{\phi}^{(i)}_a$, respectively.
We shall confirm below, see section \ref{fstpert}, that these deformations do not break the $\cw_\infty[0]$ algebra.

\subsection{Exactly marginal operators in the twisted sector}

The other exactly marginal operators arise from the $2$-cycle twisted sector, i.e., they are the
half-descendants of the BPS states with $h=\bar{h}=\frac{1}{2}$. From the coset viewpoint, the ground
state of the relevant twisted sector transforms in the representation \cite{Gaberdiel:2014cha}
\bal
&\Biggl(\Big[\frac{k}{2},0,0,\ldots,0\Bigr];\Bigl[\frac{k}{2}+1,0,0,\ldots,0\Bigr];k+(N+2)\Biggr)
\ .\label{twgr3}
\eal
However, for the purpose of doing the actual perturbation calculation, it is more convenient
to describe them directly in terms of the symmetric orbifold language. 

For definiteness let us consider the $2$-cycle twisted sector corresponding to the permutation $(12)$. Then the
free fields with $a\geq 3$ behave as before, while out of the two fields associated to $a=1,2$, we can form
the antisymmetric and symmetric combination
\begin{equation}
P^{(i){\rm A}}=\frac{1}{\sqrt{2}}(P^{(i)}_1-P^{(i)}_2)\label{zeroferm}\ , \qquad 
P^{(i) {\rm S}}=\frac{1}{\sqrt{2}}(P^{(i)}_1+P^{(i)}_2)\ ,
\end{equation}
where $P$ stands for one of $\psi$, $\bar\psi$, $\partial\phi$ or $\partial\bar\phi$. The $P^{(i) {\rm S}}$
fields have modes with the usual mode numbers (integers for bosons, half-integer for fermions in the NS-sector),
while for the $P^{(i){\rm A}}$ fields, the moding is reversed, i.e., integer for the fermions in the NS-sector, and
half-integer for the bosons. The non-trivial OPEs between these combinations are  (for the case of the left-moving fermions)
\alp[\psi^{(i) {\rm A}}(z) \, \bar\psi^{(j) {\rm A}}(w)&\sim \frac{\delta^{ij}}{z-w}+ {\cal O}(1) \\
\psi^{(i) {\rm S}}(z)\, \bar\psi^{(j) {\rm S}}(w)&\sim \frac{\delta^{ij}}{z-w}+ {\cal O}(1) \ .
] 
Note that under the $(12)$ permutation action, the symmetric combination is invariant, while the anti-symmetric combination
picks up a sign. Thus the states that survive the orbifold projection involve an even number of odd generators (both for
left- and right-movers). This condition will then also guarantee that the surviving states satisfy 
$h-\bar{h}\in\mathbb{Z}$. 

\subsubsection{The perturbing fields}

Next we need to introduce some notation for the twisted sector ground states. Let us denote the ground state of the twisted 
sector that is characterised by 
\alp[ \psi^{(1) {\rm A}}_0 \, |\Psi^0\>=0 \ ,\qquad \psi^{(2)  {\rm A}}_0\, |\Psi^0\>=0  \label{vacvan}] 
as $ |\Psi^0\>$. 
Then by applying the fermionic zero modes we define the states
\begin{equation}
|\Psi^+\>=\bar{\psi}^{(1){\rm A}} _0\, |\Psi^0\> \ , \qquad |\Psi^-\>=\bar{\psi}^{(2) {\rm A}}_0\, |\Psi^0\>\ , \qquad 
|\Psi^3\>=\bar{\psi}^{(1){\rm A}}_0\, \bar{\psi}^{(2){\rm A}}_0\, |\Psi^0\> \ . 
\end{equation}
The $|\Psi^\pm\>$  transform as a doublet under the $\mathfrak{su}(2)_-$ algebra, while the $|\Psi^0\>$ and $|\Psi^3\>$ are singlets. 
The actual perturbing states are then the super-descendants of the BPS gound states $|\Psi^\pm\>$.  There are only two non-vanishing, 
independent descendants, which we denote as 
\begin{eqnarray}
|\F^+\> &= & \bigl(-{\a}^{(1){\rm A}}_{-\frac{1}{2}}\bar{\psi}^{(1){\rm A}}_0
\bar{\psi}^{(2){\rm A}}_0+\bar{\a}^{(2){\rm A}}_{-\frac{1}{2}}\bigr)\, |\Psi^0\> \label{pert+} \\[4pt]
|\F^-\>&=& \bigl(\bar{\a}^{(1){\rm A}}_{-\frac{1}{2}}+{\a}^{(2){\rm A}}_{-\frac{1}{2}}\bar{\psi}^{(1){\rm A}}_0
\bar{\psi}^{(2){\rm A}}_0\bigr)\, |\Psi^0\>\ . \label{pert-}
\end{eqnarray}
With the above normalisation conventions, these states have unit norm and are orthogonal to one another. 
The $\pm$ label can be identified with the charge under the global $\mathfrak{su}(2)$ algebra. Obviously, there 
are similar anti-chiral states (that have to be combined with these), and in total there are therefore $4$ exactly
marginal deformations. 
However, only one combination of the $4$ deformations preserves the global ${\rm SO}(4)$ symmetry;
this is the exactly marginal operator that corresponds to switching on the string tension. For the computations
in this paper, the precise form of this combination is, however, not important since the right-moving perturbing field
only enters rather trivially, and the effect of the perturbation seems to be independent of which of the two left-moving
super-descendants $|\F^\pm\>$  of the ground state are considered. We have done most of the following computations 
for the case of $|\F^+\>$; we have also checked in some cases (in particular for the quadratic operators up to spin $s\leq 5$) that 
the perturbation by $|\F^-\> $ leads exactly to the same conclusion.

\section{First order deformation analysis}\label{fstpert}

With these preparations we can now study the behaviour of the chiral fields under the perturbation by the exactly marginal field 
$|\F\>$ from above. (As was mentioned there, we have mainly done the analysis for $|\F\> \equiv  |\F^+\>$.)
To first order, i.e., considering the $3$-point function that involves two chiral fields and one perturbing field, the answer is always trivial.
This is simply a consequence of the fact that the perturbing field has $h=\bar{h}=1$, and since the other two fields have $\bar{h}=0$,
the anti-chiral part of the correlation function vanishes, see also the discussion in \cite{Gaberdiel:2013jpa}. 

Nevertheless, there is a `first-order' analysis that determines whether a given chiral field will pick up an anomalous dimension. 
The criterion for the spin $s$ field $W^{(s)}$ of the chiral algebra to remain chiral is that, see e.g., \cite{Gaberdiel:2013jpa}
\bal
\cn(W^{(s)})\equiv \sum_{l=0}^{\lfloor s+h_\F \rfloor-1}\frac{(-1)^l}{l!} \, (L_{-1})^l W^{(s)}_{-s+1+l}\, \F=0\ ,\label{defmaster}
\eal
where $\F$ is the perturbing state. Note that we have not assumed that $\F$ is primary, and 
hence the sum in \eqref{defmaster} runs over a slightly larger index set than in \cite{Gaberdiel:2013jpa}. Actually, 
$\cn(W^{(s)})$ has the interpretation \cite{Cardy:1989da,Fredenhagen:2007rx}
\begin{equation}\label{pert1}
\partial_{\bar{z}} W^{(s)} (z,\bar{z}) = g\, \pi\,  \cn(W^{(s)}) \ ,
\end{equation}
where $g$ is the coupling constant.

\subsection{Deformation by the untwisted sector perturbation fields}

Let us begin by studying the perturbation by the untwisted sector fields \eqref{upert}. 
The chiral fields at $s=\frac{1}{2}$, see eq.~(\ref{4f}), are purely fermionic and are not affected by the 
bosonic perturbations \eqref{upert}. The same statement applies to all the 
spin $s=1$ fields, except those described by eq.~(\ref{bs1}). For the latter, the corresponding zero modes 
vanish on $\F$, and the $+1$ modes map $\F$ to the vacuum, which in turn is annihilated by $L_{-1}$. Hence, 
we conclude that ${\cal N}=0$ in (\ref{defmaster}), and thus also all spin $s=1$ fields are not lifted by this perturbation. 
We should mention that a similar conclusion was also reached in \cite{Gaberdiel:2014yla}, where the perturbation by the field
$({\rm f};\bar{\rm f})$ was studied. For generic $k$ this perturbation breaks the spin $s=1$ symmetry, see
eq.~(5.4) of \cite{Gaberdiel:2014yla}, but the effect disappears in the $k\rightarrow \infty$ limit (in which the
$({\rm f};\bar{\rm f})$ field is closely related to the above deformations). 

We can carry out a similar analysis for the generators of higher spin, and we find that, as expected, 
the perturbation by the untwisted sector fields \eqref{upert} does not break any of the 
$\cw_\infty[0]$ generators, nor those of the stringy extension.

\subsection{Deformation by the twisted sector perturbation fields}

For the case of the perturbation by a twisted sector field, the situation is more interesting.
By construction, the perturbing fields (\ref{pert+}) and (\ref{pert-}) are $\mathfrak{su}(2)_-$ singlets,
\begin{equation}
 J^{\alpha}_0\, |\F^\pm\>=0  \qquad \alpha \in\{ 3, \pm\} \ .
\end{equation}
Thus it follows that the perturbation by $\F=\F^\pm$ preserves the $\mathfrak{su}(2)_-$ symmetry. On the other
hand, for the spin $1$ current from the first non-trivial ${\cal N}=4$ multiplet, we find 
\alp[
\nn \cn(V^{(1)0}) & =(V^{(1)0})_0 |\F^+\> \\
&=\Bigl(\bar{\psi}^{(1){\rm A}}_0{\psi}^{(1) {\rm A}}_0
+\bar{\psi}^{(2){\rm A}}_0{\psi}^{(2){\rm A}}_0 - 1\Bigr)\,
\Bigl(-{\a}^{(1){\rm A}}_{-\frac{1}{2}} \, \bar{\psi}^{(1){\rm A}}_0\, \bar{\psi}^{(2){\rm A}}_0
+\bar{\a}^{(2){\rm A}}_{-\frac{1}{2}}\Bigr)\, 
|\Psi^0\>\\
\nn &=\Bigl(-{\a}^{(1){\rm A}}_{-\frac{1}{2}}\bar{\psi}^{(1){\rm A}}_0\,
\bar{\psi}^{(2){\rm A}}_0-\bar{\a}^{(2){\rm A}}_{-\frac{1}{2}}\Bigr)\, |\Psi^0\>\ .]
The resulting state does not vanish, and hence the spin $1$ field $V^{(1)0}$ is not conserved by the perturbation.
A similar computation can also be done for the other generators of the extended chiral algebra with a similar result.
Thus it appears that the perturbation by $\F^+$ breaks the higher spin symmetry down to the ${\cal N}=4$ superconformal
algebra. This is what one should expect for the perturbation by the field that switches on the tension. A similar conclusion 
applies also to $\F^-$.

\section{Second order deformation analysis}\label{sndpert}

The analysis of the previous section has given some evidence for the fact that the perturbation by $\F=\F^\pm$ breaks
the higher spin symmetry down to the ${\cal N}=4$ superconformal algebra. In this section we want to calculate 
the relevant anomalous dimensions quantitatively; this will allow us to determine the masses of the corresponding
fields, and hence enable us to see whether these fields lie on different Regge trajectories. 
In order to do this computation, we now need to perform a second order analysis.

Let us begin by explaining the general structure of our computation. We consider 
the normalised perturbed two point functions
\alp[
\<W^{(s)i}(z_1)W^{(s)j}(z_2)\>_\F&=\frac{\<W^{(s)i}(z_1)W^{(s)j}(z_2)e^{\Delta S}\>}{\<e^{\Delta S}\>}\ , \quad 
\Delta S=g \int d^2 w \, \F(w,\bar{w})\ ,
]
where the coupling constant $g$ is dimensionless. Expanding in powers of $g$, we get 
\alp[\nn
\qquad &\<W^{(s)i}(z_1)W^{(s)j}(z_2)\>_\F-\<W^{(s)i}(z_1)W^{(s)j}(z_2)\>\\
&\qquad =\frac{g^2}{2} \, \Big(\int d^2 w_1 \, d^2 w_2\,  \<W^{(s)i}(z_1)\, W^{(s)j}(z_1)\, \F(w_1,\bar{w}_1)\, \F(w_2,\bar{w}_2)\> 
\label{genpert}\\
&\qquad \qquad-\int d^2 w_1 \, d^2 w_2 \, \<W^{(s)i}(z_1)\, W^{(s)j}(z_2)\>\, \, 
\<\F(w_1,\bar{w}_1)\, \F(w_2,\bar{w}_2)\>\Big)+\co(g^3) \ , \nonumber 
] 
where the $\co(g)$ term vanishes as explained at the beginning of the previous section, and hence the leading correction to the 2-pt function 
appears at second order. To read off the anomalous dimension from this calculation we note that the $2$-point function of any
quasiprimary operator is of the form 
\alp[\<W^{(s)i}(z_1)W^{(s)j}(z_1)\>_\F\sim \frac{c^{ij}}{(z_1-z_2)^{2(s+\gamma^{ij})}(\bar{z}_1-\bar{z}_2)^{2\bar{\gamma}^{ij}}}\ ,
\label{WWpert}]
which can be expanded, for small $\gamma^{ij}$, as 
\alp[ \<W^{(s)i}(z_1)W^{(s)j}(z_1)\>_\F\sim\frac{c^{ij}}{(z_1-z_2)^{2s}}\,
\bigl(1-2\g^{ij}\ln(z_1-z_2)-2\bar{\g}^{ij}\ln(\bar{z}_1-\bar{z}_2)+\cdots \bigr)\ ,]
where $\gamma^{ij}=\bar{\gamma}^{ij}$, because of locality.
Thus we can read off the anomalous dimension $\gamma^{ij}$ from the $\log$-term in the perturbed $2$-point function. In general,
however, we need to `diagonalise' the perturbed $2$-point functions as typically different fields (of the same unperturbed conformal
dimension) will mix at this order. In order to simplify the analysis we shall assume that the fields $W^{(s)i}$ for $i=1,\ldots,N(s)$
form an orthonormal basis of the spin $s$ fields in the unperturbed theory, i.e., that 
\begin{equation}
\langle W^{(s)i}(z_1) \, W^{(s)j}(z_2) \rangle = \delta^{ij}\, (z_1 - z_2)^{-2 h_i}  \ .
\end{equation}
Then the relevant mixing matrix that we have to diagonalise 
turns out to equal (see Appendix~\ref{evintegral}, where we present two different approaches to this calculation)
\alp[\g^{ij}=\frac{g^2\pi^2 \big\< [\cn(W^{(s)i})](z_1)\,  [(\cn(W^{(s)j})](z_2)\big\>}
{ (z_1 - z_2)^{h_i +h_j}}
\ ,\label{gamma}]
where $\cn(W)$ was defined in (\ref{defmaster}).
Writing out the definition of $\cn(W)$, this can be further simplified to 
\alp[\g^{ij}= g^2\pi^2 \sum_{m=1-s}^{s\bmod 1 }(-1)^{\lceil s\rceil -1-\lfloor m\rfloor}\binom{2s-2}{s-1-m}
\<\F|{W}^{(s)i}_{-m}W^{(s)j}_m|\F\> \ .\label{gammaev}]
\smallskip

One can also arrive at the same conclusion following the analysis in \cite{Aharony:2006hz}. To this end we observe that,
because of locality, the anomalous dimension must be the same for the left- and right-moving conformal dimension. 
Then, since $\bar{L}_{-1} = \partial_{\bar{z}}$, the anomalous dimension matrix is proportional to 
\begin{equation}
\langle \partial_{\bar{z}} W^{(s)i}|  \partial_{\bar{z}} W^{(s)j}\rangle 
= 2 \, \langle W^{(s)i} | \bar{L}_0 | W^{(s)j} \rangle \ , 
\end{equation}
where we have used that $\bar{L}_{-1}^\dagger = \bar{L}_{1}$, as well as the fact that $\bar{L}_{1} W^{(s)j} = 0$. 
On the other hand, using (\ref{pert1}), the left-hand-side can be described in terms of the $\cn(W^{(s)i})$, and thus
the matrix elements of $\bar{L}_0$ in the orthonormal basis $W^{(s)k}$ equal the matrix elements of 
$\cn(W^{(s)k})$.\footnote{We thank Wei Li for a discussion about this point.}

\subsection{A more structural approach}

We can also understand some of the entries of the mixing matrix using the ${\cal W}$-algebra representation theory. In order
to explain this, let us consider the case where $W^{(s)}$ and $W^{(t)}$ are ${\cal W}$-highest weight states of spin $s$, and 
$t$, respectively. Then a certain $W^{(s)}$-descendant' of $W^{(t)}$
\bal
W^{s+t-1}=\sum_{m=0} \,\frac{(-1)^m(2s-2)_{(m)}}{m!\, (2s+2t-4)_{(m)}} (L_{-1})^m W^{(s)}_{m+1-s} W^{(t)}\, 
\equiv \cn(W^{(s)})_t W^{(t)} \label{cmpopt}
\eal
defines a quasiprimary field of spin $s+t-1$. For the calculation of the mixing matrix, we need to determine the associated
vector of eq.~(\ref{defmaster}), i.e., 
\bal
\nn \cn(W^{s+t-1})&=\sum_{p'=0}^{s+t-1} \frac{(-1)^{p'}}{p'!} (L_{-1})^{p'} W^{s+t-1}_{-(s+t-1)+1+p'}\F\\
\nn &=\frac{1}{{2s+2t-4 \choose 2s-2}}\sum_{p'=0}^{s+t-1} \sum_{k=0}^{2s-2} {p' \choose k} {2s+2t-4-p'\choose 2s-2-k} \frac{(-1)^{p'}}{p'!} \\
& \qquad \qquad \times \  (L_{-1})^{p'} [W^{(s)}_{1-s+k},W^{(t)}_{-t+1+p'-k}]\F\ .
\eal
After a slightly tedious computation, we find that the resulting vector can be written as 
\begin{equation}
\cn(W^{s+t-1})=\cn(W^{(s)})_{t}\, \cn(W^{(t)})_{1}\, \Phi -\cn(W^{(t)})_{s}\, \cn(W^{(s)})_{1}\, \Phi \ , \label{nnn}
\end{equation}
where $\cn({W^{(s)}})_t$ was already implicitly defined in (\ref{cmpopt}), and is explicitly given as 
\begin{equation}\label{5.12}
\cn({W^{(s)}})_t=\sum_{m=0} \,\frac{(-1)^m(2s-2)_{(m)}}{m!\, (2s+2t-4)_{(m)}} (L_{-1})^m W^{(s)}_{m+1-s}\ .
\end{equation}
(Note that $\cn({W^{(s)}})_1\, \Phi = \cn(W^{(s)})$, see eq.~(\ref{defmaster}).)
This formula leads to constraints among the entries of the mixing matrix; for example, it shows directly that 
if both $W^{(s)}$ and $W^{(t)}$ do not acquire an anomalous dimension, i.e.,
$\cn(W^{(s)}) = \cn(W^{(t)})=0$,  the same is true for the $W^{(s)}$-descendant 
$W^{s+t-1}$ of $W^{(t)}$.

\subsection{Relation to bulk masses}

Ultimately, we are not directly interested in the anomalous dimensions of the higher spin fields, but in the associated
bulk masses. It is well known that the bulk mass of an arbitrary $p$-form field is related to the conformal dimension on 
the boundary via, see, e.g., \cite{Aharony:1999ti}
\begin{equation}  
\Delta=\frac{1}{2}(2+2\sqrt{(1-p)^2+m^2})\ .
\end{equation}
Solving for the mass $m$ leads to the equation 
\alp[m^2=(\Delta-1)^2-(p-1)^2\ .] 
In particular, if $\Delta=p$, then $m^2=0$. Now, if the spin $s$ operator with $(h,\bar{h})=(s,0)$ and $\Delta=h+\bar{h}=s$ 
acquires an anomalous dimension, its bulk mass becomes 
\alp[m^2&=
(s+\g(s)-1)^2-(s-1)^2\\
\nn &=\g(s)(2s+\g(s)-2)\ \approx \  2\g(s)(s-1)\ ,]
where we have used the approximation that the anomalous dimension is much smaller than the spin. 
A standard flat space Regge trajectory would therefore correspond to the situation where the anomalous dimension
$\gamma(s)$ is a constant, independent of $s$. As we will see below, it seems that for the case at hand, the 
anomalous dimension behaves as $\gamma(s) \sim \log s$, at least without taking the mixing between the different
fields into account.

\section{Explicit anomalous dimensions}\label{sec:exp}

Now we have accumulated all the ingredients to derive explicit expressions for the anomalous dimensions. Before we 
get started with describing the explicit  results, there are a few general features of the results we should point out.

\subsection{The structure of the analysis} \label{mix}

In general, the matrix $\gamma^{ij}$ will not be diagonal, and in order to extract the anomalous dimensions from the calculation of 
(\ref{gammaev}), we need to diagonalise it. As we have reviewed in section~\ref{sec:hsf}, the spin fields 
can be organised in columns labeled by the number of the underlying free fields
$m=2,3,\ldots$,  see eq.~(\ref{hs}). The fields that appear in the $m$'th column have spin $s\geq m-2$, and since
to leading order only fields of the same conformal dimension have non-trivial matrix elements $\gamma^{ij}$, for any
fixed spin, the spin $s$ part of the matrix $\gamma^{ij}$ is finite-dimensional.

There are some selection rules that guarantee that not all the different spin fields can mix with one another. First of all,
since all the free fields carry non-trivial $\mathfrak{u}(1)$ charge, it is easy to see that mixing can only take place
between fields from the ${\cal W}_\infty$ representations $(0;[n,0,\ldots,0,\bar{n}])$ for which $n-\bar{n}$ has the same value. 
As a consequence, only columns of order $m$ (with $m=n+\bar{n}$) that differ by an even number, can mix.
Furthermore, since the perturbation does not break the ${\cal N}=4$
superconformal algebra, the anomalous dimensions must be the same for each member of a given ${\cal N}=4$ multiplet --- we
have also checked this explicitly for a few cases --- and indeed only fields that sit in the same ${\cal N}=4$ multiplets can mix. 
Because of this, we shall always give the anomalous dimension of full ${\cal N}=4$ multiplets, and we shall label
them by the spin of the lowest field (and the appropriate $\mathfrak{su}(2)_+\oplus \mathfrak{su}(2)_-$ labels).

However, even taking both of these considerations
into account, the problem of diagonalising the complete mixing matrix becomes quickly very complicated, and we have
only solved it completely for rather low spin, see Section~\ref{sec:lowspin} below. 
For higher values of $s$, we have only performed the diagonalisation within the subset of bilinear 
and trilinear fields, see Section~\ref{sec:partialdiag}. 

\subsection{The anomalous dimensions at low spin}\label{sec:lowspin} 

In this section we give our explicit results for small spin. Let us consider the different values in turn.

\subsubsection{Spin $1$}

At spin $1$ we have the lowest multiplet of the original higher spin algebra, i.e., $R^{(1)}$ from
(\ref{Winf}). In addition, we have a conjugate pair of $R^{(1)}$ representations coming from 
(\ref{addq}). Because of $\mathfrak{u}(1)$ charge conservation, these fields cannot mix, and
hence the relevant $\gamma^{ij}$ matrix is already automatically diagonal. All its diagonal
entries turn out to equal
\be\label{dhs1}
\delta h \bigl(R^{(1)} \bigr)_{m=2} = \frac{1}{2} \, \frac{g^2\pi^2}{N+1} \ .
\ee

\subsubsection{Spin $\frac{3}{2}$}

At spin  $\frac{3}{2}$, the only muliplets arise from the $(0;[2,0,\ldots,0,1])$ and $(0;[1,0,\ldots,0,2])$ sectors,
see  eq.~(\ref{2.17}). Since these two multiplets cannot mix (because they have different $\mathfrak{u}(1)$ charge),
the $\gamma^{ij}$ matrix is again diagonal, and we can read off the anomalous dimensions from the diagonal
entries. They turn out to equal
\be
\delta h \bigl(R^{(3/2)} \bigr)_{m=3} = \frac{3}{4}\, \frac{g^2\pi^2}{N+1}  \ .
\ee

\subsubsection{Spin $2$}\label{exs2}

The situation at spin $2$ is more interesting. There is one $R^{(2)}$ multiplet from the original higher spin algebra, see
eq.~(\ref{Winf}). It can mix with the $R^{(2)}$ multiplet from the $m=4$ term in the $(0;[2,0,\ldots,0,2])$ representation,
see eq.~(\ref{2.19}). The mixing matrix turns out to be 
\be
\gamma^{ij}(R^{(2)} ) = \frac{g^2\pi^2}{N+1} \, \left( \begin{matrix}
\frac{9}{16}   &    \frac{3}{16}\sqrt{6} \cr
\frac{3}{16}\sqrt{6}   &   \frac{15}{8}
\end{matrix}
\right) \ ,
\ee
and, after diagonalisation, it leads to the two anomalous dimensions
\be\label{dhs2}
\delta h \bigl(R^{(2)} \bigr)_{m=2} = 0.418 \, \frac{g^2\pi^2}{N+1} \ , \qquad \delta h \bigl(R^{(2)} \bigr)_{m=4} = 2.020 \, \frac{g^2\pi^2}{N+1} \ ,
\ee
where we have made the assignment $m=2$ and $m=4$ since the relevant eigenvectors are predominantly
from the $m=2$ and $m=4$ part, respectively. (The relevant prefactors are relatively close to
$\frac{9}{16} = 0.562$ and $\frac{15}{8}=1.875$.)

In addition, there is one pair of $R^{(2)}({\bf 2},{\bf 1})$ multiplets in the $(0;[3,0,\ldots,0])$ and $(0;[0,\ldots,0,3])$ representations,
see eq.~(\ref{2.13}). Again these states cannot mix since they have opposite $\mathfrak{u}(1)$ charge, and the 
anomalous dimensions turn out to equal
\be
\delta h \bigl(R^{(2)}({\bf 2},{\bf 1}) \bigr)_{m=3} = \frac{3}{2} \, \frac{g^2\pi^2}{N+1} \ .
\ee
Similarly, a pair of $R^{(2)}({\bf 2},{\bf 1})$ multiplets sit in the $(0;[1,0,\ldots,0,2])$ representation and its conjugate, 
$(0;[2,0,\ldots,0,1])$, see eq.~(\ref{2.17}), and their anomalous dimensions also turn out to equal
\be
\delta h \bigl(R^{(2)}({\bf 2},{\bf 1}) \bigr)_{m=3} = \frac{3}{2} \, \frac{g^2\pi^2}{N+1} \ .
\ee

\subsubsection{Spin $\frac{5}{2}$}

At higher spin we have only evaluated the anomalous dimensions for some states. For example, at 
spin $s=5/2$, the multiplet $R^{(5/2)}({\bf 2},{\bf 2})$  from the $(0;[2,0,…,0,2])$ representation at $m=4$ 
does not participate in mixing with other multiplets, and its anomalous dimension equals
\begin{equation}
\delta h \Bigl(R^{(5/2)}({\bf 2},{\bf 2}) \Bigr)  =\frac{5}{2}\, \frac{g^2\pi^2}{N+1} \ .
\end{equation}

\subsubsection{Spin $3$}\label{exs3}

We have also worked out some examples at spin 3. First, of all there is mixing between the $R^{(3)}$ multiplet of the 
original higher spin algebra and the $R^{(3)}$ multiplet in the $(0;[2,0,\ldots,0,2])$ representation at $m=4$. The mixing matrix 
is of the form
\be
\gamma^{ij}(R^{(3)})=  \frac{g^2\pi^2}{N+1} \, \left( \begin{matrix}
\frac{21}{32}   &    \frac{55\sqrt{5}}{512\sqrt{2}} \cr
\frac{55\sqrt{5}}{512\sqrt{2}}   &   \frac{2435}{768}
\end{matrix}
\right) \ ,
\ee
and the anomalous dimensions are, upon the diagonalisation of the mixing matrix,
\be
\delta h \bigl(R^{(3)} \bigr)_{m=2} =  0.645 \, \frac{g^2\pi^2}{N+1} \ , \qquad 
\delta h \bigl(R^{(3)} \bigr)_{m=4} = 3.182 \, \frac{g^2\pi^2}{N+1} \ .
\ee
There is furthermore one $R^{(3)}({\bf 2},{\bf 1})$ multiplet at $m=3$ term in the 
$(0;[2,0,\ldots,0,1])$ representation, see eq.~(\ref{2.17}). It can mix with the $R^{(3)}({\bf 2},{\bf 1})$ multiplet from the 
$(0;[3,0,\ldots,0,2])$ representation at $m=5$. The mixing matrix turns out to be 
\be
\gamma^{ij}(R^{(3)}({\bf 2},{\bf 1}) ) = \frac{g^2\pi^2}{N+1}\, \left( \begin{matrix}
\frac{51}{32}   &    \frac{9}{8}\sqrt{2} \cr
\frac{9}{8}\sqrt{2}   &   \frac{21}{4}
\end{matrix}
\right) \ ,
\ee
and, after diagonalisation, it leads to the two anomalous dimensions
\be\label{dhs3}
\delta h \bigl(R^{(3)}({\bf 2},{\bf 1})  \bigr)_{m=3} = 0.998 \, \frac{g^2\pi^2}{N+1} \ , \qquad 
\delta h \bigl(R^{(3)}({\bf 2},{\bf 1}) \bigr)_{m=5} = 5.845 \, \frac{g^2\pi^2}{N+1} \ ,
\ee
where we have made the assignment as in the previous case. There is also a multiplet $R^{(3)}({\bf 3},{\bf 1})$ coming from 
the $(0;[2,0,\ldots,0,2])$ representation at $m=4$, see eq.~(\ref{2.19}); it cannot mix with any other multiplet, and its anomalous
dimension therefore equals the diagonal entry which is 
\be
\delta h \bigl(R^{(3)}({\bf 3},{\bf 1}) \bigr)_{m=4} = \frac{15}{4} \, \frac{g^2\pi^2}{N+1} \ .
\ee

\subsection{Partial diagonalisation for higher spin}\label{sec:partialdiag}

Unfortunately, the size of the mixing matrix increases rather quickly with the spin, and the calculation becomes soon very complicated. 
In particular, it is therefore not feasible to determine the large spin behaviour of the anomalous dimensions in this manner. In order
to obtain some idea of the functional form, we have resorted to studying the partial diagonalisation problem
where we only diagonalise the mixing matrix among the fields of a given value of $m$. In particular, for the `leading' 
$m=2$ family this is rather simple since no mixing can take place: the original higher spin fields (\ref{Winf}) and the 
fields from (\ref{addq}) cannot mix among each other since they have different $\mathfrak{u}(1)$ charge, and in each
family, there is only one multiplet of a given spin. Thus this $\gamma^{ij}$ submatrix is automatically diagonal, and it is enough 
to determine the diagonal matrix elements. This can be done in closed form with the result
\begin{equation}
\g^{(s)} =\frac{g^2\pi^2 \sum_{p=0}^{s}(-1)^{s-p}\binom{2s}{s-p}\,P_2(s,p)}{(N+1)\, E_2(s)}\ ,\label{anagamma}
\end{equation}
where
\bal   E_2(s)=&\sum_{q=0}^{s-1} \sum_{p=0}^{s-1}(-1)^{s+1+p+q} \binom{s}{q}{s\choose q+1}  
\binom{s}{p}{s\choose p+1} \\
\nn &\qquad \times  \bigl((-2)_{(q)}(-2-q)_{(s-p-1)}(-2)_{(s-q-1)}(q-s-1)_{(p)} \bigr)\ ,\\
P_2(s,p)=&\sum_{n=3/2}^{p-3/2} n (p-n)f(s,p,n)f(s,-p,n-p)\\
\nn & + \tfrac{3}{2}\, (-1)^{s+1} \, \Q(p-2)f(s,p,1/2)f(s,-p,-1/2)\,(p-1/2)\\
\nn & +\tfrac{1}{2}\,\delta_{p,1}\, f(s,1,1/2)f(s,-1,-1/2)\ ,\\[4pt]
f(s,p,n)=&\sum_{q=0}^{s-1} (-1)^q \binom{s}{q}{s\choose q+1} \,(-1-p+n)_{(s-q-1)}\,(-1-n)_{(q)}\ , \label{eq:6.7}
\eal
and $s$ is the spin of the higher spin current multiplet. Here $(a)_{(n)}$ is defined by 
\begin{equation}
(a)_{(n)}=a(a-1)\cdots (a-n+1) \ .
\end{equation}
This formula was derived for the multiplets of spin $s$ arising from (\ref{Winf}); for the multiplets 
from (\ref{addq}), the relevant formula looks different (see the ancillary file), but gives rise to exactly the same
values for $\gamma^{(s)}$ (for odd spin $s$). 
For the multiplets up to $s\leq 5$, we have worked out the anomalous dimensions separately for each 
member of the multiplet (and confirmed that it is indeed the same for all the fields of a given multiplet); for $s\geq 6$,
the analysis was done for a specific state in the $({\bf 3},{\bf 1})$ representation at $h=s+1$. 

\begin{figure}[bht]
\centering
\includegraphics[width=1\textwidth]{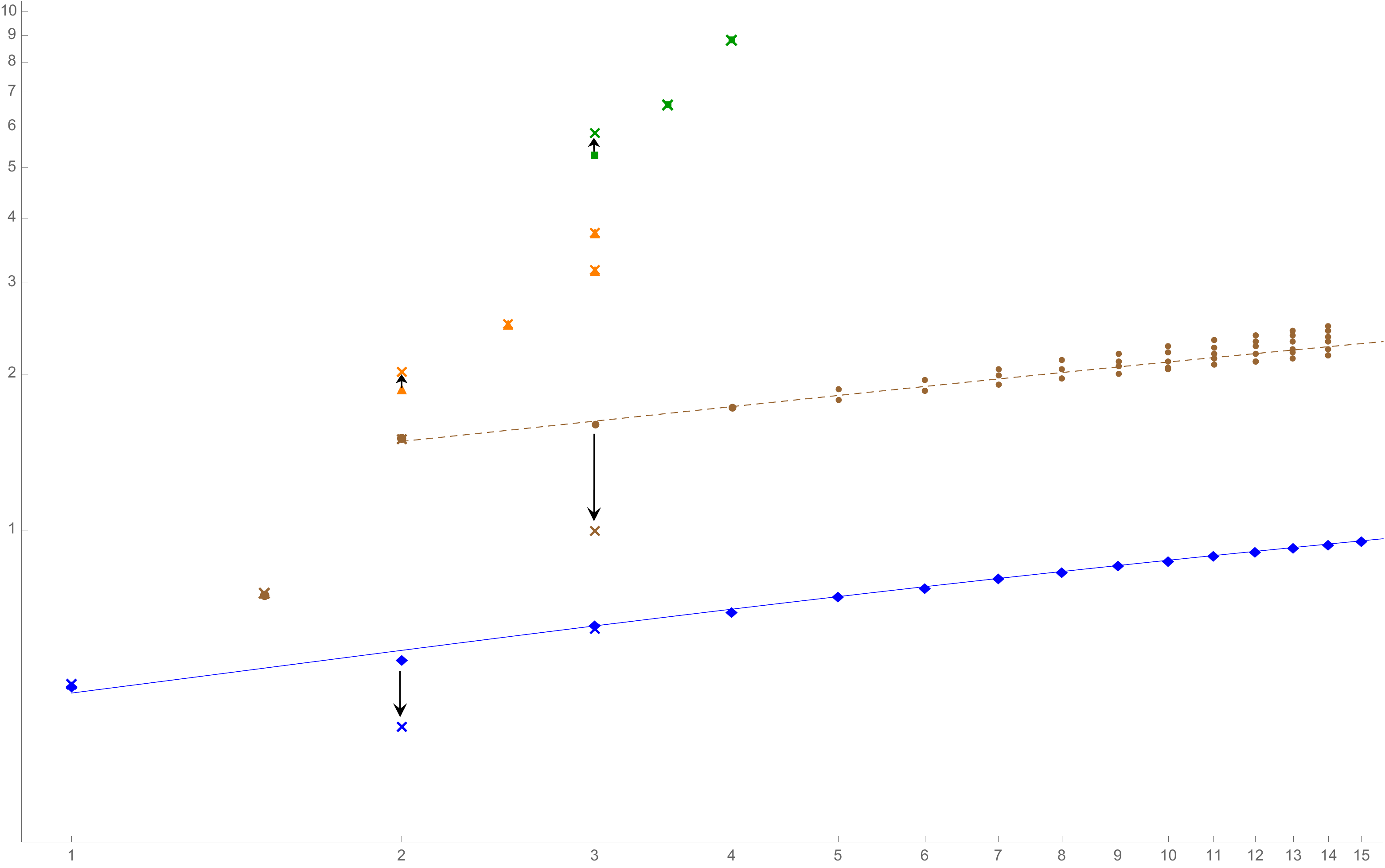}
\caption{The anomalous dimensions, in units of $\p^2g^2/(N+1)$, as a function of the spin in a log-log plot.
Dots represent the diagonal values (e.g., for $m=2$), or the eigenvalues after partial diagonalisation among 
the fields of the corresponding value of $m$ (e.g., for $m=3$), while crosses describe the actual eigenvalues
after complete diagonalisation (with arrows indicating where the relevant eigenvalue originates from). 
The solid blue and dashed brown lines are the fitting curves (\protect\ref{coshfit}) and (\protect\ref{cubfit}), respectively.
We have used the colour code blue diamonds ($m=2$); brown circles ($m=3$); orange triangles ($m=4$); and green squares ($m=5$).
}
\label{fittri}
\end{figure}

While we have not been able to simplify the formula for $\g^{(s)}$ for general $s$, we
have plotted the result, see the solid blue curve in the log-log plot of Figure~\ref{fittri}.  
The resulting curve can be fitted quite well by the function 
\alp[\g^{(s)}_2=0.20293\,  \log \bigl(7.04703\, s + 3.84921\bigr)\ .\label{coshfit}]
\medskip

We have also done a similar analysis for the fields with $m=3$. While the fields from the different sectors 
$(0;[n,0,\ldots,0,\bar{n}])$ with $n+\bar{n}=3$ cannot mix (because of $\mathfrak{u}(1)$ charge conservation), 
non-trivial mixing will occur due to the multiplicities $n(s)>1$ and $m(s)>1$ in eqs.~(\ref{ns}) and (\ref{ms}), respectively.
We have solved this mixing problem explicitly up to spin $s\leq 14$, and we have plotted these partially diagonalised eigenvalues
by the brown dots in Figure~\ref{fittri}. 
For larger spin, we have again only worked out the diagonal matrix elements for one family of fields 
that transforms in the $({\bf 4},{\bf 1})$ representation; the curve that fits these diagonal anomalous dimensions is 
\begin{equation}\label{cubfit}
\gamma_3^{(s)} = 0.476933  \log \bigl(7.61932\, s + 14.7676\bigr) \ ,
\end{equation}
and it is also included (as the dashed brown curve) in the log-log plot of Figure~\ref{fittri}.  In this figure we have also included 
the exact eigenvalues of some low-lying fields (involving up to $m=5$), some of which were explicitly described in 
section~\ref{sec:lowspin}; in those cases, where there was a genuine mixing, we have also indicated with an arrow
the shift in eigenvalue due to the mixing.

\section{Discussion}\label{sec:discuss}

In the previous section we have studied the anomalous dimensions of the higher spin currents of the symmetric orbifold
theory as one turns on the perturbation that corresponds to the string tension. While at the tensionless point, the higher spin fields
of the Vasiliev theory --- these correspond to the fields of the original ${\cal W}_\infty$ algebra --- form a decoupled
subsector, these fields begin to couple with the other stringy symmetry generators once the perturbation is switched on.
As a consequence, the complete perturbation analysis is quite complicated. However, we have managed to solve it exactly for the first
few values of the spin; we have also managed to obtain good analytic control over the diagonal entries of the mixing
matrix, at least for the original higher spin generators --- these are quadratic in the underlying free fields --- and the 
generators that are cubic in the free fields. Our results are summarised in Figure~\ref{fittri}.

The results we have obtained seem to be very nicely in line with the idea that the original higher spin generators correspond
to the `leading' Regge trajectory, i.e., have the lowest mass (or anomalous dimension) for a given spin. We should
stress that once mixing occurs (as will generically be the case), the actual eigenvectors will be linear combinations
of fields from the different families (parametrised by $m$), see e.g., the explicit calculation for the $s=2$ field
in Section~\ref{exs2}. However, it is legitimate to continue labelling the eigenvalues (and the associated eigenvectors)
by the family from which they predominantly arise. In particular, the lowest eigenvalue will be associated to the family
with the smallest diagonal entry in the mixing matrix. Our explicit calculations --- we have also worked out some of the 
diagonal entries of the mixing matrix for the $m=4$ and $m=5$ families --- suggest that the lowest eigenvalue of a given spin
comes from the quadratic fields, i.e., that the extended ${\cal W}_\infty$ algebra (where we include also the generators
arising from $(0;[2,0,\ldots,0])$ and $(0;[0,\ldots,0,2])$) describes the `leading Regge trajectory'.

By the same token, the cubic generators seem to describe the first sub-leading Regge trajectory. Furthermore, 
the dispersion relation, at least before any mixing is taken into account,
seems to be of the form $E = s + a \log s$ (for some suitable value of $a$), which is what 
one should have expected for an AdS background.\footnote{We should mention that in higher dimensions, there
are no additional symmetry generators beyond those of the Vasiliev theory in the large $N$ limit, and hence no mixing can occur. The
corresponding calculations (including those for AdS$_3$) therefore only determined the `diagonal' matrix elements,
without any diagonalisation.}
In particular, a dispersion relation of this kind was found, 
in a series of papers \cite{Gubser:2002tv,Frolov:2002av,Roiban:2007jf,Roiban:2007dq,Giombi:2009gd}, 
for slow-rotating long strings on AdS$_5\times {\rm S}^5$, 
and the result was matched to a perturbative field theory calculation in \cite{Belitsky:2006en,Eden:2006rx,Beisert:2006ez},
see \cite{Beisert:2010jr} for a review. In the slow-rotating regime, the angular momentum on ${\rm S}^5$ is negligible, and
the result also directly applies to the AdS$_3$ background that is of relevance to us. Actually, this conclusion is only valid for 
the AdS$_3$ case with pure RR flux; for pure NSNS flux, it was shown in \cite{Loewy:2002gf} that $\gamma(s)$ is
instead constant, while for mixed NSNS/RR flux, the anomalous dimension also contains a $(\log(s))^2$ contribution
\cite{David:2014qta}. We have also tried to fit our leading trajectory with an extra $(\log(s))^2$ term, and found the fitting 
coefficient of this term  to be seven orders of magnitude smaller than the coefficient of the $\log(s)$ term. Thus our 
analysis gives credence to the belief that the symmetric orbifold theory is dual to string theory on AdS$_3$ with pure RR flux. 

The computations we have done in this paper apply fairly directly also to the symmetric orbifold of K3 \cite{k3}, since the 
generators of the chiral algebra in that case form a subset of our generators, 
and the exactly marginal operators are the same as the ones we have used in our computation.
\smallskip

It would be very interesting to push this calculation further, and in particular solve the complete mixing problem
up to some higher spins. This would allow us to determine the shape of the actual dispersion relations along the different
Regge trajectories. It would also be very interesting to re-derive these anomalous dimensions
from the dual AdS viewpoint; for the case of the ${\cal N}=3$ higher spin theory of \cite{Creutzig:2014ula} 
this was done for one spin $2$ field in \cite{Hikida:2015nfa}. It would be
interesting to see if this calculation can be extended to higher spins and applied in our context.

\section*{Note Added}

Some time after this paper was posted on the arXiv, \cite{AndrejStepanchuk:2015wsq} and \cite{Banerjee:2015qeq} appeared
which claim that the dispersion relation of spiky strings in mixed-flux backgrounds does not contain the $(\log (s))^2$ term that is 
present in the folded string analysis of \cite{David:2014qta}. 
If the dispersion relation of \cite{AndrejStepanchuk:2015wsq,Banerjee:2015qeq} is indeed the one relevant for our problem, 
we cannot deduce from our analysis that the background in question has to have pure R-R flux; we can only arrive at the weaker conclusion that the 
background cannot have pure NS-NS flux. (This is because, for a pure NS-NS background the dispersion relation would be linear, see 
for example eq.~(6.0.8) of \cite{AndrejStepanchuk:2015wsq} in the limit $q\rightarrow 1$.) 
We thank Justin David, Igor Klebanov and Arkady Tseytlin for communications about this issue.

\section*{Acknowledgements}

We thank Niklas Beisert, Nikolay Bobev, Henriette Elvang, Rajesh Gopakumar, Matthew Headrick, Finn Larsen, Albion Lawrence 
and Wei Li for useful conversations.  
M.R.G.\ is grateful to Durham University for hospitality during the final stages of this work. 
C.P.\ acknowledges the hospitality from the Korea institute for Advanced Study, University of Michigan, Ann Arbor and Brandeis University at 
various stages of this work. The research of C.P.\ is supported by a grant of the Swiss National Science Foundation, while I.G.Z.\ is 
supported in part by the National Science Foundation under CAREER Grant No.\ PHY10-53842, and in part by the Department of Energy 
under Grant No.\ DE-SC0009987. This research was finally partly supported by the NCCR SwissMAP, funded by the Swiss National Science 
Foundation.

\appendix

\section{Free field realisation of the chiral fields}\label{ffall}

Following on from the analysis of section~\ref{sec:hsf}, we describe in this appendix the fields at spin 
$s={\frac{3}{2}}$, and also present the free field realisation of the stress-energy tensor $T$. Dropping for ease of notation the sum over the copies, we have
\bal
&G^{++}=-2 (:{\partial\phi}^{(2)}\bar{\psi}^{(1)}:+:\partial\bar{\phi}^{(1)}{\psi}^{(2)}:) \,, \,\quad 
G'^{++}=2 (-:{\partial\phi}^{(2)}\bar{\psi}^{(1)}:+:\partial\bar{\phi}^{(1)}{\psi}^{(2)}:)\,, \nn \\ 
& G^{--}=2 (:{\partial\phi}^{(1)}\bar{\psi}^{(2)}:+:\partial\bar{\phi}^{(2)}{\psi}^{(1)}:) \,, \quad ~~~ 
G'^{--}=2 (:{\partial\phi}^{(1)}\bar{\psi}^{(2)}:-:\partial\bar{\phi}^{(2)}{\psi}^{(1)}:)\,, \nn \\
&  G^{-+}=2 (:{\partial\phi}^{(1)}\bar{\psi}^{(1)}:-
:\partial\bar{\phi}^{(2)}{\psi}^{(2)}:) \,, \quad ~~~ 
G'^{-+}=2 (:{\partial\phi}^{(1)}\bar{\psi}^{(1)}:+:\partial\bar{\phi}^{(2)}{\psi}^{(2)}:)\,, \nn \\
&  G^{+-}=2 (-:{\partial\phi}^{(2)}\bar{\psi}^{(2)}:+
:\partial\bar{\phi}^{(1)}{\psi}^{(1)}:)\,, \quad  
G'^{+-}=-2 (:{\partial\phi}^{(2)}\bar{\psi}^{(2)}:+:\partial\bar{\phi}^{(1)}{\psi}^{(1)}:)\ ,  
\eal
all of which are elements of the ${\cal W}_\infty[0]$ algebra. The additional generators are the fields
\bal
& C_{21} =\sum_{a=1}^{N+1}  :\psi^{(i)}_a\,\psi^{(j)}_a\,\bar{\psi}^{(k)}_a: \ , \qquad
& C_{12} = \sum_{a=1}^{N+1}  :\psi_a^{(i)}\,\bar{\psi}^{(j)}_a\,\bar{\psi}^{(k)}_a:\ , \\
& C_{20}= \sum_{a=1}^{N+1}  :{\partial\phi}^{(1)}_a{\psi}^{(2)}_a:\ , \qquad  & 
C_{02}=\sum_{a=1}^{N+1} :\partial\bar{\phi}^{(i)}_a\bar{\psi}^{(j)}_a: \ . 
\eal
The stress-energy tensor is of the form
\bea\label{T-freefield}
&& T =
\sum_{a=1}^{N+1} \bigg\{ (:\partial\phi^{(1)}_a\bar\partial\phi^{(1)}_a:+:\partial\phi^{(2)}_a\bar\partial\phi^{(2)}_a:)+\\
&&\qquad\qquad +\,
\frac{1}{2} \, 
(:\partial\bar\psi^{(1)}_a\psi^{(1)}_a:-:\bar\psi^{(1)}_a\partial\psi^{(1)}_a:)
+\frac{1}{2}\,(:\partial\bar\psi^{(2)}_a\psi^{(2)}_a:-:\bar\psi^{(2)}_a\partial\psi^{(2)}_a:)\bigg\}\ .\nn
\eea
For the ${\cal W}_\infty$ operators with higher spin we can work out the free field realisations by taking recursively 
OPEs of the above generators.

\section{The second order analysis} \label{evintegral}

There are at least two natural approaches to calculate the integral (\ref{genpert}), and we shall 
sketch them in the following.

\subsection{Using Stokes' theorem} 

In this approach, the first step is to use the OPEs of the currents to rewrite the integrand of the first term on the right hand of 
(\ref{genpert}) 
\alp[
\nn &\sum_{m,n}(z_1-w_1)^{-h_1-m}(z_2-w_2)^{-h_2-n}\<0|\big(W^1_m\F^{1}(w_1,\bar{w}_1)\big)\big(W^2_n\F^{2}(w_2,\bar{w}_2)\big)|0\>\\
\nn &\quad +\sum_{m,n}(z_1-w_2)^{-h_1-m}(z_2-w_1)^{-h_2-n}\<0|\big(W^1_m\F^{2}(w_2,\bar{w}_2)\big)\big(W^2_n\F^{1}(w_1,\bar{w}_1)\big)|0\>\\
\nn &\quad +\sum_{m,n}(z_1-w_1)^{-h_1-m}(z_2-w_1)^{-h_2-n}\<0|\big(W^1_m W^2_n\F^{1}(w_1,\bar{w}_1)\big)\F^{2}(w_2,\bar{w}_2)|0\>\\
 &\quad +\sum_{m,n}(z_1-w_1)^{-h_1-m}(z_1-w_1)^{-h_2-n}\<0|\F^{2}(w_2,\bar{w}_2)\big(W^1_m W^2_n\F^{1}(w_1,\bar{w}_1)\big)|0\>\,.\label{int1}
]
The $\bar{w}_1$ and $\bar{w}_2$ dependence is rather trivial --- it equals $\frac{1}{(\bar w_1-\bar w_2)^2}$ --- since only the
perturbing fields have a non-trivial anti-chiral dependence. Thus each term above leads to an integral of the form 
\alp[
\nn & \int d^2 w_1 \, d^2 w_2 \, (x_1-w_1)^{-p} \, (x_2-w_2)^{-q} \, \frac{\<\co^1(w_1)\co^2(w_2)\>}{(\bar w_1-\bar w_2)^2}\\
&\qquad =\int d^2 w_1\,  (x_1-w_1)^{-p}\int d^2 w_2 \, \frac{\pa_{\bar{w}_2} (\frac{\<\co^1(w_1)\co^2(w_2)\>}{(\bar w_1-\bar w_2)})}{(x_2-w_2)^q}
\label{Stokes}\\
\nn &\qquad =\int d^2 w_1 (x_1-w_1)^{-p}\frac{1}{2i}\, \oint_{x_2} d w_2\,  \frac{(\frac{\<\co^1(w_1)\co^2(w_2)\>}{(\bar w_1-\bar w_2)})}{(x_2-w_2)^q} \ , ]
where we have used integration by parts. Now the contour integral can be evaluated by standard methods, and it yields
\alp[
\nn &=\frac{(-1)^q\pi}{(q-1)!}\,  \int d^2 w_1\,  \frac{(x_1-w_1)^{-p}}{(\bar w_1-\bar x_2)}\, 
\<\co^1(w_1)\big((L_{-1})^{q-1} \co^2\big)(x_2)\> \\
\nn &=\frac{(-1)^q\pi}{(q-1)!} \int d^2 w_1 \frac{\pa_{\bar{w}_1} \big(\<\co^1(w_1)\big((L_{-1})^{q-1}
\co^2\big)(x_2)\>\log(|\bar w_1-\bar x_2|^2)\big)}{(x_1-w_1)^p}\\
\nn &=\frac{(-1)^q\pi}{(q-1)! 2i} \oint_{x_1} d w_1 \frac{\<\co^1(w_1)\big((L_{-1})^{q-1}
\co^2\big)(x_2)\>\log(|\bar w_1-\bar x_2|^2)}{(x_1-w_1)^p}\\
 &=\frac{(-1)^{p+q}\pi^2}{(q-1)! (p-1)!}  {\<\big((L_{-1})^{p-1}\co^1\big)(x_1)\big((L_{-1})^{q-1}
\co^2\big)(x_2)\>\log(|\bar x_1-\bar x_2|^2) \ , }
 \label{2ptint}
]
where, in the second line, we have used the identity 
\bal \frac{1}{\bar{z}}=\pa_{\bar{z}} \log(z\bar{z}) 
\ . \label{holodof}
\eal 
Note that the appearance of the factor $z\bar{z}$ is required to make the function single-valued. 
Applying this result to the expression \eqref{int1} with $\F^1=\F^2$,\fn{The last two lines do not give rise to 
logarithmic terms since either $p$ or $q$ is zero; they therefore do not play a role for our computation.}
we find for the coefficient of $\p^2 \log(|\bar z_1-\bar z_2|^2)=2\p^2 \log(|\bar z_1-\bar z_2|)$ the expression
\alp[
\nn &2\sum_{m,n}
\frac{(-1)^{h_1+m+h_2+n-2}}{(h_2+n-1)! (h_1+m-1)!} \<\big((L_{-1})^{h_1+m-1}W^1_m\F^{}\big)(z_1)\big((L_{-1})^{h_2+n-1}
W^2_n\F^{}\big)(z_2)\>\\
 &\quad =2 \,\big\<\big(\cn(W^1)_1\F\big)(z_1)\,\big(\cn(W^2)_1\F\big )(z_2)\big\>\ ,
\label{int2}
] 
where we have used the definition of (\ref{5.12}), see also \eqref{defmaster}, 
and the $\cn$ here is the corresponding vertex operators of the states. 
Including the overall normalisation factor then leads directly to eq.~(\ref{gamma}). 

In order to obtain from this eq.~(\ref{gammaev}), we recall that the $2$-point function is related to the inner product
of the associated state as 
\be\label{main2pt}
\langle V(\psi_2,z_2) V(\psi_1,z_1) \rangle = (-1)^{h_1}\,\<e^{-L_1} \psi_2|e^{-L_1}\psi_1\>\,  (z_2 - z_1)^{-h_1 - h_2} \ ,
\ee
which simplifies, for the case that both fields are quasi-primary, to the more familiar relation
\be\label{main2ptprime}
\langle V(\psi_2,z_2) V(\psi_1,z_1) \rangle = (-1)^{h_1}\,\<\psi_2|\psi_1\>\,  (z_2 - z_1)^{-h_1 - h_2} \ .
\ee
A simple computation reveals that $L_1 \,\cn(W^s)_1\,\F=0$, and thus the $2$-point function in \eqref{gamma} is, up to an
irrelevant factor of $(-1)^s$, the norm of the two  $\cn(W^s)_1\F$ states.

To simplify the evaluation of the norm, we now move the $L_1$'s in the bra-vector $\<\cn(W^i,\F)|$ to the right, where
they annihilate the ket-vector  $|\cn(W^s)_1\F\rangle$; similarly we can move the $L_{-1}$ generators of $|\cn(W^s)_1\F\rangle$
to the left. After a slightly tedious computation we then find 
\alp[ & \<\cn({W}^{(s)i},\F)\, \, \cn(W^{(s)j},\F)\> \\ 
& \qquad =(-1)^{s-1}\sum_{m=1-s}^{s\bmod 1 }(-1)^{\lceil s\rceil -1-\lfloor m\rfloor}\binom{2s-2}{s-1-m}\<\F|{W}^{(s)i}_{-m}\,W^{(s)j}_m|\F\>\ .\label{normsimp1}]
This is then eq.~(\ref{gammaev}).

\subsection{Using separation of variables }

The second approach differs from the above in the treatment of the integral (\ref{Stokes}). The idea of the computation
is to split off from the double integral the UV divergence, so that the remainder is UV finite (and contains the anomalous dimension).
More concretely, we use the M\"obius symmetry to rewrite the $4$-point function in the integrand as 
\alp[
\nn &\<W^i\Bigl(z_1+\frac{\e(z_1-z_4)(z_1-z_2)}{(z_4-z_2)} \Bigr)\,\F(z_2)\, \F(z_3)\, W^j(z_4)\>\\
&=(z_4-z_1)^{-2 s_i} \big| \frac{(z_1-z_4)^2}{(z_3-z_1)^2(z_2-z_4)^2} \big|^2 \<W^i(\epsilon^{-1} ) \, \F(1)\, \F(x)\, W^j(0)\>\ ,
]
where now the integration variables are $z_2$ and $x$. Because of the change of variables, there is a Jacobian factor 
\alp[
|dz_2|^2|d z_3|^2=|dz_2|^2 \, |dx|^2 \, \Big|\frac{\pa (z_2,z_3)}{\pa (z_2,x)}\Big|^2\ ,
]
and the original integral becomes
\alp[
&\int \int |dz_2|^2 \, |d z_3|^2 \, \<W^i(z_1+\e)\, \F(z_2)\, \F(z_3)\, W^j(z_4)\>\\
&=\int \, |dz_2|^2\, |dx|^2 \, (z_4-z_1)^{-2 s_i} \big| \frac{(z_1-z_4)}{(z_1-z_2)(z_2-z_4)} \big|^2 
\<W^i(\epsilon^{-1} )\, \F(1)\, \F(x)\, W^j(0)\> \ .
]
Carrying out the $z_2$ integration, we now get
\alp[
& 2\p \log (\L^2 |z_4-z_1|^2)\,  (z_4-z_1)^{-2 s_i} \int |dx|^2\, \<W^i(\e^{-1} )\, \F(1)\F(x)W^j(0)\>\ ,\label{4pt1}
]
where $\L$ is a UV cutoff. It remains to do the $dx$ integral. Since the left- and right-moving parts of the correlation function 
decouple, we can rewrite (\ref{4pt1}) as 
\bea\label{tildeF-sep}
&&\Big\langle W^{i}(\e^{-1})\, \Phi(1,1)\, \Phi(x,\bar x)\, W^{j}(0)\Big\rangle=
\frac{1}{(1-\bar x)^2}\,\Big\langle W^{i}(\e^{-1})\,\Phi(1)\,\Phi(x)\, W^{j}(0)\Big\rangle \ .\nonumber
\eea
As in \eqref{2ptint} above, this integral now becomes
\bea\label{int-stokes}
&&\qquad\qquad \oint_{C}\frac{dx}{2i}\,\frac{1}{(1-\bar x)}\,\Big\langle W^{i}(\e^{-1})\, \Phi(1)\, \Phi(x)\, W^{j}(0)\Big\rangle\ ,
\eea
where $C$ is a contour that circles around all the insertion points. The integrand reads
\bal
\nn &\Big\langle W^{i}(\e^{-1})\, \Phi(1)\, \Phi(x)\, W^{j}(0)\Big\rangle
 =\Big\< 0\Big| W^i_{h_i}\,\Phi(1)\;\Phi(x) \,W^j_{-h_i}\Big|0 \Big\>\\
\nn &\qquad =\Big\<0\Big|  [W^i_{h_i}\,,\Phi(1)]\;\Phi(x) \,W^j_{-h_i}\Big|0\Big\>
+\Big\<0\Big|  \Phi(1)\;[W^i_{h_i}\,,\Phi(x)] \,W^j_{-h_i}\Big|0\Big\> \\
\nn & \qquad \quad +\Big\< 0\Big| \Phi(1)\;\Phi(x) \, [W^i_{h_i}\,,W^j_{-h_i}]\Big|0 \Big\>\ ,
\eal
where the third term is the disconnected part of the $4$-point function that is subtracted out in the second line of 
\eqref{genpert}. The first two terms can be evaluated using, e.g., the techniques of \cite{Gaberdiel:1993mt}
\bal
\nn &=\Bigl\<0\Big|  \sum_{m=1-h_i}^{h_i}\binom{2h_i-1}{m+h_i-1} (W^i_{m } \Phi)(1)\;\Phi(x) \,W^j_{-h_i}\Big|0 \Bigr\>\\
&\qquad +\Bigl\<0\Big|  \Phi(1)\; \sum_{p=1-h_i}^{h_i}\binom{2h_i-1}{m+h_i-1}x^{h_i-m} (W^i_{m } \Phi)\,W^j_{-h_i}\Big|0\Bigr\>\ .
\eal
Repeating the same step then finally leads to 
\bal
\nn &=\sum_{m=1-h_i}^{h_i}\sum_{n=1-h_j}^\infty\binom{2h_i-1}{m+h_i-1} (-1)^{n+h_j-1}\frac{(-1)^{1}}{(1-x)^{1-m-n+1}}\, 
\< e^{-L_1}    W^j_{n} W^i_{m } \Phi\big|\Phi\>\\
\nn  &\qquad 
+ \sum_{m=1-h_i}^{h_i}\sum_{n=1-h_j}^\infty \binom{h_j+n-1}{h_j-h_j}(-1)^{n+h_j-1}(x)^{-n-h_j} \binom{2h_i-1}{m+h_i-1} \\
\nn & \qquad \qquad \qquad \frac{(-1)^{1-n} }{(1-x)^{1-m+1-n}} \, 
 \< e^{-L_1} W^i_{m } \Phi\big| e^{-L_1} W^j_n\Phi \> \ ,
\eal
where we have used \eqref{main2pt}. In the first line, the only $x$-dependence appears in the $1-x$ pole which will not
contribute to the contour integral. Thus only the second and third line contribute, and the contour integral 
(\ref{int-stokes}) becomes
\alp[
\nn &\oint \frac{dx}{2i} \sum_{m=1-h_i}^{h_i}\sum_{n=1-h_j}^\infty \binom{h_j+n-1}{h_j-h_j}\binom{2h_i-1}{m+h_i-1}
\frac{(-1)^{h_j}}{(x)^{n+h_j}} \frac{ \< e^{-L_1} W^i_{m } \Phi\big| e^{-L_1} W^j_n\Phi \>}{(1-x)^{1-m+1-n}}\\
&\quad =\p \sum_{m=1-h_i}^{h_i}\sum_{n=1-h_j}^\infty \binom{2h_i-1}{m+h_i-1}\binom{-m+h_j}{n+h_j-1}(-1)^{h_j} 
{ \< e^{-L_1} W^i_{m } \Phi\big| e^{-L_1} W^j_n\Phi \>}\ .\label{toev}
]
Rearranging the $L_{\pm 1}$ as in the previous subsection 
\bal
\nn & \big\< e^{- L_1}W^i_{m}\Phi\, \big| e^{- L_1}W^j_n\Phi\big\> \\
& \qquad = \sum_{p=0}^{-\lfloor m\rfloor} (-1)^{m-n}\binom{{h_i-1-m}}{p}\binom{h_j-1-n}{p+m-n}
\big\< W^i_{p+m}\Phi\, \big|   W^j_{p+m}\Phi\big\>\ ,
\eal
eq.~\eqref{toev} can be rewritten as 
\bal
& = {(-1)^{s-1}}\sum_{m=1-s}^{s\bmod 1 }(-1)^{\lceil s\rceil -1-\lfloor m\rfloor}\binom{2s-2}{s-1-m}\<\F|{W}^{(s)i}_{-m}W^{(s)j}_m|\F\>\,,\label{cmpsimp1}
\eal
which agrees precisely with eq.~\eqref{normsimp1}.
\fn{In the case when the currents are fermions, we have picked the branch of the square-root (due to half-integer spin) such 
that the resulting anomalous dimensions have the same sign as the anomalous dimensions of the bosons in the same supermultiplet. 
Notice that there is no such difficulty in the computation of the anomalous dimensions of the bosonic fields.}

\bibliographystyle{utphys}
\bibliography{versionarxiv}

\end{document}